\documentclass[aip,pop,amsmath,amssymb,showpacs,reprint,floatfix,lengthcheck]{revtex4-1}
\usepackage[colorlinks,linkcolor=blue,citecolor=blue,bookmarks,pdfstartview=FitH]{hyperref}
\usepackage{graphicx}
\usepackage{dcolumn}
\usepackage{bm}
\usepackage{url}
\usepackage{amssymb}
\usepackage{enumerate}
\usepackage{enumitem}

\begin{document}

\preprint{AIP/123-QED}

\title{\textcolor{blue}{Parametric upconversion of lower hybrid wave by runaway electrons in tokamak}}

\author{A. Kuley}
\email{animesh47@gmail.com}
\affiliation{Department of Physics,Indian Institute of Technology Delhi, New Delhi-110016, India.}
\author{V. K. Tripathi}
\affiliation{Department of Physics,Indian Institute of Technology Delhi, New Delhi-110016, India.}
\date{\today}

\begin{abstract}
A  kinetic formalism of parametric decay of a large amplitude lower hybrid pump wave into runaway electron mode and a uppersideband mode is investigated. The pump and the sideband exert a ponderomotive force on runaway electrons, driving the runaway mode. The density perturbation associated with the latter beats with the oscillatory velocity due to the pump to produce the sideband. The finite parallel velocity spread of the runaway electrons turns the parametric instability into a stimulated compton scattering process where growth rate scales as the square of the pump amplitude. The large phase velocity waves thus generated can potentially generate relativistic electrons.

\end{abstract}

\maketitle

\section{Introduction}
Relativistic runaway electrons are potentially a serious problem in advanced tokamaks like International Thermonuclear Experimental Reactor (ITER) \cite{helander2002runaway,shimada2007overview}. Runaway electrons (REs) with energies upto several mega-electron-volts have been observed during numerous disruption in large tokamaks, such as Joint European Torus (JET) \cite{wesson1989disruptions,gill2002behaviour,riccardo2003disruptions,plyusnin2006study}, Frascati Tokamak Upgrade (FTU) \cite{PhysRevLett.97.165002,martin2004runaway}, Japan Torus (JT-60U) \cite{yoshino2000runaway,kawano2002disruption}, Hefei Tokamak-7 (HT-7) \cite{chen2006dynamics,chen2007enhancement}. They are a matter of serious concern as they can cause severe damage to the first wall structure on impact. Runaway electrons can also arise in smaller numbers under normal tokamak operation, especially during startup and in low density plasmas. Two mechanisms are primarily attributed to runaway generation : primary generation through Dreicer accleration \cite{PhysRev.117.329}, and secondary generation provided by the avalanche effect \cite{rosenbluth1997theory}. For ITER disruptions, it has been predicated that avalanching would dominate and turn as much as two thirds of the predisruption current into the runaway current \cite{PhysRevLett.92.205004}.

Current carrying fast electrons are generated/sustained by lower hybrid waves through parallel electron Landau damping when the Cerenkov resonance condition is fulfilled. Recently Martin-Solis \textit{et al}.\cite{PhysRevLett.97.165002} observed experimentally,  large production of runaway electrons $($upto $\sim$ 80 percent of the predisruption plasma current$)$ during a disruptive termination of discharge heated with lower-hybrid waves in FTU. Chen \textit{et al}.\cite{chen2006dynamics} investigated the effect of lower hybrid waves on runaway production in the HT-7 tokamak, and showed that the presence of lower hybrid waves can greatly enhance the runaway production with high residual electric field.  In Ref.\cite{PhysRevLett.38.162} Liu and Mok proposed an elegant theory for the nonlinear evolution of runaway electron distribution and time dependent synchrotron emission from tokamak. 

In this paper we study the parametric upconvesrsion of a lower hybrid wave into a low frequency electrostatic mode and a lower hybrid upper sideband. The low frequency mode is in resonant interaction with the runaway electrons. The lower hybrid pump wave $(\omega_{0}, \textbf{k}_{0})$, imparts an oscillatory velocity to electrons. When this is combined with a lower frequency density perturbation, a nonlinear current is produced driving the sideband $(\omega_{2}, \textbf{k}_{2})$. The sideband and the pump exert a ponderomotive force on electrons, driving the low frequency perturbation.

The paper is organized as follows : in sec. II we obtain the runaway electron susceptibility. Sec. III contains the nonlinear coupling, and growth rate have been calculated in sec. Electron acceleration by the deacay wave have been discussed in sec IV. Discussions have been given in sec. V.

\section{Runaway electron susceptibility}
We model the tokamak by a uniform plasma of background electron density $n_{0}^{0}$ in shearless magnetic field $\textbf{B}=B_{0} \hat{z}$. The plasma has a component of runaway electrons of density $n_{0r}^0$. The initial runaway electron distribution, before the wave-particle interaction, is determined from the kinetic equation with the boundary condition that satisfies the avalanche growth rate of the runaway density $dn_{r}/dt = n_{r}(E-1)/(c_{z}\tau_{c})$ \cite{rosenbluth1997theory}, giving 
\begin{equation}
 n_{r}=n_{r0}exp[(E-1)t/(\tau_{c}c_{z})],
\end{equation}
where $E=eE_{\parallel}\tau_{c}/m_{e0}c$ is the normalized parallel electric field, assumed to be constant in time, $c$ is the speed of light, $\tau_{c}=4\pi\epsilon_{0}^2 m_{e0}^2 c^3/n_{e}e^4 ln \Lambda$ is the collision time for relativistic electrons, $c_{z}=\sqrt{3(Z+5)/\pi}$ln$\Lambda$, Z is the effective ion charge and $n_{r0}$ is the seed produced by primary generation. Thus $n_{0r}^0$ is the value of runaway density $n_{r}$ at the time when lower hybrid wave is launched. In a tokamak disruption, this initial distribution function of the relativistic tail of runaway electrons is \cite{pokol2008quasi,fulop2006destabilization}
\begin{equation}
 f_{0}(p_{z},p_{\perp})=\frac{n_{r0}^{0} \alpha}{2\pi c_{z}p_{z}}exp\biggl(-\frac{p_{z}}{c_{z}}-\frac{\alpha p_{\perp}^2}{2p_{z}}\biggr),
\end{equation}
where $\textbf{p}=\gamma \textbf{v}/c$ is the normalized relativistic momentum of the runaway electrons and $\alpha=(E-1)/(Z+1)$, with $\gamma$ as the relativistic factor.

We perturb this equilibrium by an elecrostatic perturbation
\begin{equation}
 \phi=A e^{-i(\omega t-\textbf{k}\cdot\textbf{r})}
\end{equation}
where $\omega\ll\omega_{c}$, $\omega_{c}=eB_{0}/m_{e0}$. The response of runaway electrons to it is governed by the Vlasov equation,
 \begin{equation}
 \frac{\partial}{\partial t}f+\textbf{v}\cdot\nabla f= -\frac{e}{m_{e0}c}\biggl[\nabla\phi-\textbf{v}\times \textbf{B}_{s}\biggr]\cdot \nabla_{\textbf{p}} f
\end{equation}
Writing $f=f_{0}+f_{1}$ and linearizing Eq.(4) we obtain
\begin{equation}
 f_{1}=-\frac{e}{m_{e0}c}\int^{t}_{-\infty} \biggl[\nabla\phi \cdot(\frac{\partial}{\partial \textbf{p}}f_{0})\biggr]_{t{'}}    dt^{'},
\end{equation}
where the integration is over the unperturb trajectory.
 Eq.(5) simplifies to give 
\begin{eqnarray}
f_{1}=-\frac{e\phi }{m_{e0}c}\frac{n_{0r}^{0} \alpha}{2\pi c_{z}p_{z}}exp\biggl(-\frac{p_{z}}{c_{z}}-\frac{\alpha p_{\perp}^2}{2p_{z}}\biggr) \sum_{l} J_{l}\biggl(\frac{k_{\perp}p_{\perp}c}{\omega_{c}}\biggr)\times\nonumber\\
\sum_{n} J_{n}\biggl(\frac{k_{\perp}p_{\perp}c}{\omega_{c}}\biggr)\biggl(\frac{k_{z}}{p_{z}}+\frac{k_{z}}{c_{z}}+\alpha\frac{l\omega_{c}}{c}-\alpha\frac{p_{\perp}^2}{2p_{z}^2}k_{z}\biggr)\times\nonumber\\
\frac{e^{i(l-n)\theta}}{\omega-\frac{k_{z} p_{z}c}{\gamma}-l\Omega},\quad\quad
\end{eqnarray}
where $J_{l}$ and $J_{n}$ are the Bessel functions of order $l$ and $n$, and $\theta$ is the gyrophase angle with $\Omega=eB/m_{e}=\omega_{c}/\gamma$. The perturbed density of  runaway electrons  turns out to be
\begin{equation}
 n_{b}=\int_0 ^{\infty} \int_0 ^{2\pi} \int_{p_{c}} ^{\infty} f_{1} p_{\perp} dp_{\perp} d\theta dp_{z}.
\end{equation}
Where $p_{c}$ is the boundary between the bulk and fast electrons (tail region) momentum space \cite{eriksson2003simulation}. For the anisotropy of the runaway distribution with relativistic electrons, with small argument of Bessel function, and by using the identity
\begin{equation}
 \int_{0}^{\infty} e^{-s^2}J_{l}^{2}(\Psi s) s ds=\frac{1}{2} I_{l}e^{-{\beta}^2/2}
\end{equation}
the perturbed density takes the form
\begin{eqnarray}
 n_{b}=\frac{n_{r} e}{m_{e0} c_{z}}\frac{k_{\perp}^2}{\omega_{c}^2\alpha}\phi\gamma\biggl[\biggl(\frac{1}{c_{z}}+\frac{k_{\perp}^2c^2}{2{\omega_{c}^2}\alpha}\biggr)\frac{e^{-bp_{c}}}{b}+\nonumber\\
 \biggl(\frac{D}{c_{z}}+\frac{k_{\perp}^2c^2}{2{\omega_{c}}^2\alpha}D
-\frac{\omega_{c}\alpha}{k_{z}c}\biggr)e^{-bD}\displaystyle\lim_{\epsilon\to 0}\int_{(p_{c}-D)b}^{\infty}\frac{q e^{-q}}{q^2+b^2 {\epsilon}^2}dq\nonumber\\
+i\pi  e^{-bD}\biggl(\frac{D}{c_{z}}+\frac{k_{\perp}^2c^2}{2{\omega_{c}}^2\alpha}D-\frac{\omega_{c}\alpha}{k_{z}c}\biggr)\biggr]\quad
\end{eqnarray}
where $b=1/c_{z}+k_{\perp}^2c^2/{\omega_{c}^2}{\alpha}$, $D=({\gamma \omega-\omega_{c}})/k_{z}c$, $q/b=p_{z}-D $.

For $(\omega \ll \omega_{c})$ we can write
\begin{equation}
 n_{b}=\frac{k^2}{e}\epsilon_{0}(\chi_{br}+i\chi_{bi})\phi, 
\end{equation}
where
\begin{eqnarray}
\chi_{br}=\frac{\omega_{pr}^2}{c_{z}}\frac{k_{\perp}^2}{\omega_{c}^2\alpha k^2}\gamma\biggl[\biggl(\frac{1}{c_{z}}+\frac{k_{\perp}^2c^2}{2{\omega_{c}^2}\alpha}\biggr)\frac{e^{-bp_{c}}}{b}+\nonumber\\
\biggl(\frac{\gamma\omega}{k_{z}cc_{z}}+\frac{k_{\perp}^2c^2}{2{\omega_{c}}^2\alpha}\frac{\gamma\omega}{k_{z}c}-\frac{\omega_{c}\alpha}{k_{z}c}\biggr)e^{-b\frac{\gamma\omega}{k_{z}c}}\times\nonumber\\
\displaystyle\lim_{\epsilon\to 0}\int_{(p_{c}-\frac{\gamma\omega}{k_{z}c})b}^{\infty}\frac{Q e^{-Q}}{Q^2+b^2 {\epsilon}^2}dQ\biggr],\nonumber\\
\chi_{bi}=\frac{\omega_{pr}^2}{c_{z}}\frac{k_{\perp}^2}{\omega_{c}^2\alpha k^2}\gamma \pi  e^{-b\frac{\gamma\omega}{k_{z}c}}\times\nonumber\\
\biggl(\frac{\gamma\omega}{k_{z}cc_{z}}+\frac{k_{\perp}^2c^2}{2{\omega_{c}}^2\alpha}\frac{\gamma\omega}{k_{z}c}-\frac{\omega_{c}\alpha}{k_{z}c}\biggr),
\end{eqnarray}
where ${\omega_{pr}}^2=n_{0r}^{0} e^2/m_{e0}\epsilon_{0}$, and $Q/b=p_{z}-\gamma\omega/k_{z}c$, $p_{c}=\sqrt{2T/m_{e0}}/c$.
 
\section{Nonlinear coupling and Growth Rate}
We consider the parametric coupling of a lower hybrid pump wave of potential 
\begin{equation}
\phi_{0}=A_{0} e^{-i(\omega_{0} t -\textbf{k}_{0\perp}\cdot\textbf{r}-k_{0z}z)}, 
\end{equation}
with a runaway electron mode
 \begin{equation}
 \phi=Ae^{-i(\omega t-\textbf{k}\cdot\textbf{r})}
\end{equation}
and an upper sideband mode
\begin{equation}
\phi_{2}=A_{2}e^{-i(\omega_{2} t-\textbf{k}_{2}\cdot\textbf{r})} 
\end{equation}
 where 
 $\omega_{2}=\omega+\omega_{0}$, and $\textbf{k}_{2}=\textbf{k}+\textbf{k}_{0}$.
The pump and sideband wave are primarily sustained by plasma electrons and ions and obey dispersion relation
\begin{equation}
 \omega_{j}^2=\omega_{LH}^2(1+\frac{k_{zj}^2}{k_{j}^2}\frac{m_{i}}{m}),
\end{equation}
where $j=0,2$, $\omega_{LH}^2=\omega_{pi}^2/(1+\omega_{p}^2/\omega_{c}^2)$, $\omega_{p}=(n_{0}^{0}e^2/m_{e0}\epsilon_{0})^{1/2}$, $\omega_{pi}=(Zn_{0}^{0}e^2/m_{i}\epsilon_{0})^{1/2}$, $m_{i}$ is the ion mass.
The $\omega$, $\textbf{k}$ mode has $\omega\backsimeq k_{z}v_{0b}^{0}$ and has prominent contribution from the runaway electrons. Here $v_{0b}^0$ is an average velocity of runaway electrons. We presume pump to have parallel phase velocity less than $v_{0b}^{0}$. Let $\omega_{0}/k_{0z}v_{0b}^{0}=\eta_{0}<1$, $\omega/k_{z}v_{0b}^{0}=1$, then $\omega_{2}/k_{2z}v_{0b}^{0}=(\omega+\omega_{0})/(k_{z}+k_{0z})v_{0b}^{0}
<1$, i.e., the upper sideband also move slower than the runaway electrons.

Had we considered lower hybrid pump wave to parallel phase velocity opposite to the velocity of the runaway electrons, $\omega_{0}/k_{0z}v_{0b}^{0}=-\eta_{0}$, we would obtain
\begin{equation}
 \omega_{2}/k_{2z}v_{0b}^{0}=\frac{\frac{\omega}{\omega_{0}}+1}{\frac{\omega}{\omega_{0}}-\frac{1}{\eta_{0}}}>1,
\end{equation}
i.e., the upper sideband moves faster than the runaways. 

The pump and the sideband waves impart oscillatory velocity to plasma electrons
\begin{eqnarray}
\textbf{v}_{j \perp}=\frac{e}{m_{e0}\omega_{c}^2}[\boldsymbol{\omega}_{c}\times\nabla_{\perp}\phi_{j}-i\omega_{j}\nabla_{\perp}\phi_{j}],\nonumber\\ 
v_{jz}=-\frac{ek_{jz}}{m_{e0}\omega_{j}}\phi_{j},
\end{eqnarray}
where j=0,2.

The oscillatory velocity of runaway electrons can be obtained from the linearized equation of motion
\begin{equation}
 m_{e0}[\frac{\partial}{\partial t}(\gamma \textbf{v}_{bj})+v_{0b}^{0}\cdot \nabla (\gamma \textbf{v}_{bj})]=e\nabla\phi_{j}-e\textbf{v}_{bj}\times \textbf{B},
\end{equation}

\begin{eqnarray}
\textbf{v}_{bj \perp}=\frac{e}{m_{e0}\omega_{c}^2}[\boldsymbol{\omega}_{c}\times\nabla_{\perp}\phi_{j}-i\gamma(\omega_{j}-k_{0z}v_{0bz}^{0})\nabla_{\perp}\phi_{j}],\nonumber\\ 
v_{bjz}=-\frac{ek_{jz}}{m_{e0}{\gamma}^3(\omega_{j}-k_{jz}v_{0bz}^{0})}\phi_{j},\nonumber\\
\end{eqnarray}

The pump and upper-sideband couples a low frequency ponderomotive force ${\textbf{F}_{P}}$ on the electrons. ${\textbf{F}_{P}}$ has two components, perpendicular and parallel to the magnetic field. The response of elctrons to  ${\textbf{F}_{P\perp}}$ is strongly supressed by the magnetic field and is usually weak. In the parallel direction, the electrons can effectively respond to ${\textbf{F}_{Pz}}$, hence, frequency nonlinearity arises at $(\omega, \textbf{k})$ mainly through ${\textbf{F}_{Pz}}=-m\textbf{v}\cdot\nabla v_{z}$.
The parallel ponderomotive force, using the complex number identity $Re \textbf{A} \times Re \textbf{B}=(1/2) Re [\textbf{A}\times \textbf{B}+ \textbf{A}^*\times \textbf{B}]$, for the background electrons can be written as 
\begin{equation}
 \textbf{F}_{pz}=ei k_{z}\phi _{p}= -(\frac{m_{e0}}{2})[\textbf{v}_{0\perp}^{*}\cdot\nabla_{\perp}v_{2z}+\textbf{v}_{2\perp}\cdot\nabla_{\perp}v_{0z}^{*}],
\end{equation}
Using Eq.(17) and considering only the dominant $\textbf{E}_{0}\times\textbf{B}$ drift terms the ponderomotive potential $\phi_{p}$ takes the form
\begin{equation}
 \phi_{p}=\frac{e\phi_{0}^*\phi_{2}}{2m_{e0}\omega_{c}^2} \frac{\textbf{k}_{2\perp}\cdot \textbf{k}_{0\perp}\times \boldsymbol{\omega}_{c}}{ik_{z}\omega_{2} \omega_{0}}[\omega k_{0z}-\omega_{0}k_{z}]
\end{equation}
The nonlinear density perturbation of the plasma electrons due to ponderomotive force can be written as
\begin{equation}
 n^{NL}=-\frac{n_{0}^{0}ek_{z}^2}{m\omega^2}\phi_{p}.
\end{equation}
The linear density perturbation due to self consistent potential $\phi$ is 
\begin{eqnarray}
 n^{L}=(k^2 \epsilon_{0}/e)\chi_{e}\phi,\nonumber\\
\chi_{e}=\frac{\omega_{p}^2}{\omega_{c}^2}\frac{k_{\perp}^2}{k^2}-\frac{\omega_{p}^2}{\omega^2}\frac{k_{z}^2}{k^2}.
\end{eqnarray}
 
For the runaway electrons ponderomotive force   can be written as
\begin{equation}
 F_{pzb}=ei k_{z}\phi _{pb}=-(\frac{m_{e0}\gamma}{2})[\textbf{v}_{b0\perp}^{*}\cdot\nabla_{\perp}v_{b2z}+\textbf{v}_{b2\perp}\cdot\nabla_{\perp}v_{b0z}^{*}]
\end{equation}

Using Eq.(19) and considering only the dominant $\textbf{E}_{0}\times\textbf{B}$ drift terms the ponderomotive potential $\phi_{pb}$ takes the form
\begin{eqnarray}
 \phi_{pb}=\frac{e\phi_{0}^*\phi_{2}}{2m_{e0}\omega_{c}^2{\gamma}^2} \frac{\textbf{k}_{2\perp}\cdot \textbf{k}_{0\perp}\times \boldsymbol{\omega}_{c}}{ik_{z}}\nonumber\\
\biggl\{\frac{k_{0z}}{(\omega_{0}-k_{0z}v_{0bz})}-\frac{k_{2z}}{(\omega_{2}-k_{2z}v_{0bz})}\biggr\}
\end{eqnarray}
One may note that ponderomotive potential is maximum when $\textbf{k}_{\perp}$ and $\textbf{k}_{0\perp}$ are perpendicular to each other. The response of runaway electrons to ponderomotive potential and the self consitent potential $\phi$
\begin{equation}
 n_{b}=\frac{k^2}{e}\epsilon_{0}(\chi_{br}+i\chi_{bi})(\phi+\phi_{pb}), 
\end{equation}
Using Eqs.(22) and (26) in the Poisson's equation $\nabla ^2 \phi = (e/\epsilon_{0}) (n+n_{b}-n_{i})$, where $n=n^{L}+n^{NL}$, we obtain
\begin{equation}
 \varepsilon \phi \cong-\chi_{e}\phi_{p}-\chi_{b}\phi_{pb},
\end{equation}
where $\varepsilon=1+\chi_{e}+\chi_{b}+\chi_{i}$.

The density perturbation at $(\omega, \textbf{k})$ couples with the oscillatory velocity of electrons, $\textbf{v}_{0}$, to produce nonlinear density perturbations at upper-sideband frequency. Solving the equation of continuity for the background elctron, 
\begin{equation}
 \frac{\partial}{\partial t}n_{2}^{NL}+\nabla(\frac{n}{2}v_{0})=0,
\end{equation}
one obtains 
\begin{equation}
 n_{2}^{NL}=\frac{n}{2\omega_{2}}(\textbf{k}_{2}\cdot\textbf{v}_{0}),
\end{equation}
and for the runaway electrons 
\begin{equation}
 n_{2b}^{NL}=\frac{n_{b}}{2\omega_{2}}(\textbf{k}_{2}\cdot\textbf{v}_{b0}).
\end{equation}
Using Eqs.(29) and (30) in the Poisson's equation for the upper-sideband wave, we obtain
\begin{equation}
\varepsilon_{2}\phi_{2}=\frac{k^2}{k_{2}^2}\frac{e\phi}{m_{e0}{\omega_{c}^2}}(1+\chi_{i})\frac{\textbf{k}_{2}\cdot{\boldsymbol{\omega}}_{c}\times \nabla_{\perp}\phi_{0}}{2\omega_{2}}
\end{equation}
where
\begin{equation}
\varepsilon_{2}=1+\frac{\omega_{p}^2}{\omega_{c}^2}+\gamma\frac{\omega_{pr}^2}{\omega_{c}^2}\frac{(\omega_{2}-k_{2z}v_{0bz})}{\omega_{2}}-\frac{\omega_{pi}^2}{\omega_{2}^2},
\end{equation}
is the dielectric function at $(\omega_{2},\textbf{k}_{2})$.

Eqs. (27) and (31) are the nonlinear coupled equations for $\phi$ and $\phi_{2}$ from which nonlinear dispersion relation can be obtained
\begin{equation}
 \varepsilon\varepsilon_{2}=\mu,
\end{equation}
where the coupling co-efficient can be written as
\begin{eqnarray}
\mu=\frac{U^2 k^2 sin^2\delta}{4}\biggl[\frac{\chi_{e}}{\omega_{2}^2}(\frac{\omega k_{0z}}{\omega_{0}k_{z}}-1)+\nonumber\\
\frac{\chi_{b}}{k_{z}{\gamma}^2\omega_{2}}(\frac{k_{0z}}{(\omega_{0}-k_{0z}v_{0bz})}-\frac{k_{2z}}{(\omega_{2}-k_{2z}v_{0bz})})\biggr],
\end{eqnarray}
with
 $U=ek_{0}\mid\phi_{0}\mid/m_{e0}\omega_{c}$ is the magnitude of $\textbf{E}_{0}\times \textbf{B}$ electron velocity, and $\delta$ is the angle between $\textbf{k}_{2\perp}$ and $\textbf{k}_{0\perp}$.
We write $\omega=\omega_{r}+i\Gamma$, $\omega_{2}=\omega_{2r}+i\Gamma$, where $\omega_{2}$ is the root of $\varepsilon_{2}$ =0. Then Eq.(33) gives the growthrate
\begin{equation}
 \Gamma=\frac{Im[\mu(1+\chi_{e}+\chi_{br}+\chi_{i}-i\chi_{bi})]}{[(1+\chi_{e}+\chi_{br}+\chi_{i})^2+\chi_{bi}^2]\frac{\partial\varepsilon_{2}}{\partial\omega_{2}}}
\end{equation}

 In order to have a numerical appreciation of results we consider the following set of parameters, corresponding to HT-7 tokamak \cite{chen2006dynamics} : background electron density $\sim$ 4$\times$ $10^{19} m^{-3}$, temperature $\sim$ 3keV, ion temperature $\sim$ 1.5 keV, magnetic field $\sim$ 2.5T, frequency of the lower hybrid pump $\sim$ 2.45 GHz and the density of the fast electron $\sim$ 2$\times$ $10^{16}$ $m^{-3}$, Z=1, ln$\Lambda$=18. 
 In Fig.1 we have plotted the normalised growthrate as function of noramlised frequency by considering the fast electron of energy $\sim$ 100 keV, for different lower hybrid pump power $U/c_{s}$=2 and 3, where $c_{s}$ is the ion sound speed, shows that the growth-rate increases significantly with the increasing of the lower hybrid power.

\begin{figure}
\centering
\includegraphics[width=0.5\textwidth ]{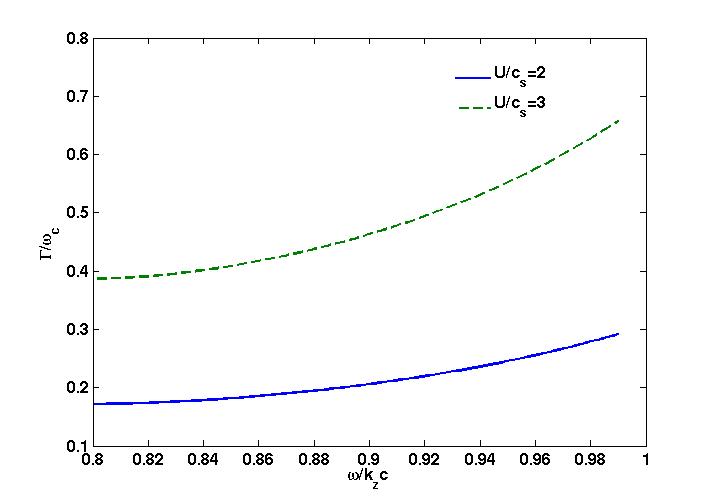}
\caption{\label{fig:epsart} Variation of normalized growthrate as a function of normalised wavenumber for fast electron of energy 100 keV.}
\end{figure}

\section{Electron Acceleration}
The dynamimcs of a runaway electron in the fast phase velocity lower hybrid sideband $\phi_{2}$ can be described by the relativistic equation of motion
\begin{equation}
 \frac{d\textbf{p}}{dt}=-e[-\nabla\phi_{2}+\textbf{v}\times \textbf{B}_{0}].
\end{equation}
Expressing $d/dt=v_{z}d/dz$ the components of Eq. (36) can be written as
 \begin{eqnarray}
  \frac{dp_{x}}{dz}=\frac{m_{e0}\gamma ek_{\perp}}{p_{z}} A_{2}sin(\omega_{2}t-k_{2x}x-k_{2z}z)-\omega_{c} m_{e0}\frac{p_{y}}{p_{z}},\nonumber\\
\frac{dp_{y}}{dz}=m_{e0}\omega_{c} \frac{p_{x}}{p_{z}},\nonumber\\
\frac{dp_{z}}{dz}=\frac{m_{e0}\gamma}{p_{z}} ek_{z}A_{2}sin(\omega_{2}t-k_{2x}x-k_{2z}z).\nonumber\\
 \end{eqnarray}
These equations are supplemented with
\begin{eqnarray}
 \frac{dx}{dz}=\frac{p_{x}}{p_{z}},\nonumber\\
\frac{dy}{dz}=\frac{p_{y}}{p_{z}},\\
\frac{dt}{dz}=\frac{m_{e0}\gamma}{p_{z}}.\nonumber
\end{eqnarray}
We introduce the dimensionless variables $P_{x}=p_{x}/m_{e0}c$, $P_{y}=p_{y}/m_{e0}c$, $P_{z}=p_{z}/m_{e0}c$, $X=x\omega_{2}/c$, $Y=y\omega_{2}/c$, $Z=z\omega_{2}/c$, $T=\omega_{2}t$, and the electron drift velocity due to upper sideband $U_{2}=ek_{2\perp}A_{2}/m_{e0}\omega_{c}$
the above Eqs. (37) and (38) reduce to
 \begin{eqnarray}
  \frac{dP_{x}}{dZ}&=&\frac{\omega_{c}U_{2}\gamma}{P_{z}c\omega_{2}} sin(T-k_{2x}X\frac{c}{\omega_{2}}-k_{2z}Z\frac{c}{\omega_{2}})- \frac{\omega_{c} P_{y}}{\omega_{2}P_{z}},\nonumber\\
\frac{dP_{y}}{dZ}&=&\frac{\omega_{c} P_{y}}{\omega_{2}P_{z}},\nonumber\\
\frac{dP_{z}}{dZ}&=&\frac{k_{2z}\omega_{c}U_{2}\gamma}{k_{2\perp}P_{z}c\omega_{2}} sin(T-k_{2x}X\frac{c}{\omega_{2}}-k_{2z}Z\frac{c}{\omega_{2}}),\nonumber\\
\frac{dX}{dZ}&=&\frac{P_{x}}{P_{z}},\nonumber\\
\frac{dY}{dZ}&=&\frac{P_{y}}{P_{z}},\nonumber\\
\frac{dT}{dZ}&=&\frac{\gamma}{P_{z}}.
 \end{eqnarray}
We solve the above equations numerically for the following set of parameters, $P_{x}(0)=0.1$, $P_{y}(0)=0$, $P_{z}(0)$=0.5 $X(0)=0.5$, $Y(0)=0$, $Z(0)=0$, and normalised pump amplitude of the upper sideband wave $U_{2}/c_{s}=2$, $\omega_{2}/k_{2z}c$ =0.99$c$, $k_{2z}/k_{2\perp}$=1/25. In Fig.2 we have plotted the electron energy, normalised to rest mass energy (on the gamma factor), as a function of distance of propagation Z. One will obtain large energy exchange between the particles and wave when phase synchronism condition is satisfied.

\begin{figure}
\centering
\includegraphics[width=0.5\textwidth ]{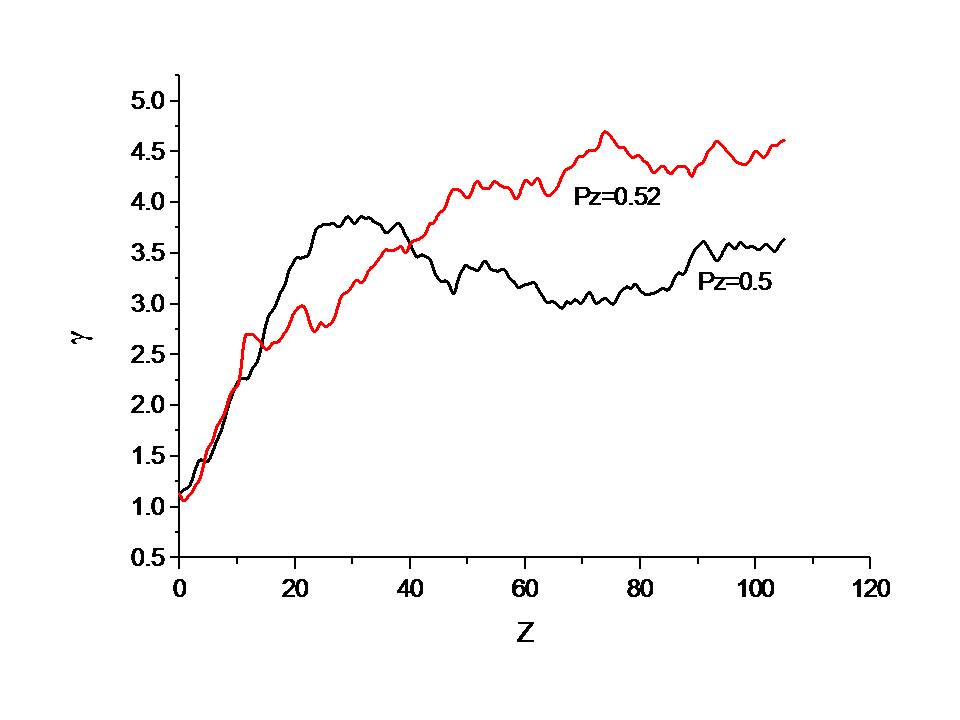}
\caption{\label{fig:epsart} Variation of electron energy normalised to rest mass energy as a function of distance of propagation for $P_{z}\sim$ 0.5 and 0.52.}
\end{figure}

\section{Discussions}

The free energy contained in the runaway electrons adds a quantum of energy to pump photon to produce the upper sideband photon. The runaway electrons open up the possibility of frequency upconversion of the lower hybrid pump wave. The parallel phase velocity of the upconverted lower hybrid wave is close to the velocity of light in vacuum. As the sideband wave acquires a large amplitude, comparable to that of the pump, it can accelerate the electrons to tens of MeV energy. For a typical case of electron drift velocity due to the upper sideband approaching 2 times the sound speed, the electron energy gain turns out to be $\sim$ 2 MeV. The maximum growth rate occur when $\delta$ is 90$^0$, where $\delta$ is the angle between $k_{2\perp}$ and $k_{0\perp}$. However the other values of $\delta$ are possible, but they will give weaker growth rate. 

The experiments on lower hybrid heating and current drive in tokamak have reported existance of MeV electrons that may be caused via parametric excitation of high parallel phase velocity waves. The energy gain by the electrons of $P_{z}\sim$ 0.5 and 0.52 with the distance of propagation upto a point, after a while the particle is taken out of the resonance, by virtue of the energy gain, and it saturates. The energy gain is primarily through the Cerenkov resonance, though the transverse filed also plays a role.

The energy gain is dependent to initial electron momentum. For given parameters, we obtain $P_{z}$=0.5-0.52, where significant energy gain occurs. For 100 KeV electrons, the Cerenkov resonance occurs when $\omega_{2}=k_{2z}v_{z}$, and for $\omega_{2}/k_{2z}$=0.99c the normalized energy gain $\gamma$ $\sim$ 5. This is the kind of energy gain we obtain in Fig. 2. 
In this paper we have ignored the toroidal and shear effects that may have profound effect on electron acceleration.

\begin{acknowledgments}
The authors are thankful to Prof. C. S. Liu of University of Maryland for his valuable discussions.
\end{acknowledgments}

\nocite{*}
\bibliography{Kuley-10}

\end{document}